\documentclass{aa}

\usepackage{amsbsy}
\usepackage{amssymb}
\usepackage{natbib}
\usepackage{txfonts}
\usepackage[dvips]{graphicx}

\newcommand{\be}{\begin{equation}}
\newcommand{\ee}{\end{equation}}
\newcommand{\beqa}{\begin{eqnarray}}
\newcommand{\eeqa}{\end{eqnarray}}

\begin{document}

\title{Probing the cosmographic parameters to distinguish between dark energy
and modified gravity models}
\author{  F. Y. Wang$^{1,2}$, Z. G. Dai$^{1}$ and Shi Qi$^{3,4}$}

\offprints{F. Y. Wang \\ \email{fayinwang@nju.edu.cn}}

\institute{Department of Astronomy, Nanjing University, Nanjing
210093, China \and Department of Astronomy, University of Texas at
Austin, Austin, TX 78712 \and Purple Mountain Observatory, Chinese
Academy of Sciences, Nanjing 210008, China \and Joint Center for
Particle, Nuclear Physics and Cosmology, Nanjing University - Purple
Mountain Observatory, Nanjing 210093, China.}

\authorrunning{F. Y. Wang et al.}
\titlerunning{Probing the cosmographic parameters}

\abstract {} {In this paper we investigate the deceleration, jerk
and snap parameters to distinguish between the dark energy and
modified gravity models by using high redshift gamma-ray bursts
(GRBs) and supernovae (SNe).} {We first derive the expressions of
deceleration, jerk and snap parameters in dark energy and modified
gravity models. In order to constrain the cosmographic parameters,
we calibrate the GRB luminosity relations without assuming any
cosmological models using SNe Ia. Then we constrain the models
(including dark energy and modified gravity models) parameters using
type Ia supernovae and gamma-ray bursts. Finally we calculate the
cosmographic parameters. GRBs can extend the redshift - distance
relation up to high redshifts, because they can be detected to high
redshifts.}{We find that the statefinder pair $(r,s)$ could not be
used to distinguish between some dark energy and modified gravity
models, but these models could be differentiated by the snap
parameter. Using the model-independent constraints on cosmographic
parameters, we conclude that the $\Lambda$CDM model is consistent
with the current data.}{}


\maketitle

\section{Introduction}
Recent observations of the Hubble relation of distant Type Ia
supernovae (SNe Ia) have provided strong evidence for acceleration
of the present universe (Riess et al. 1998; Perlmutter et al. 1999).
The observations of the spectrum of cosmic microwave background
(CMB) anisotropies (Spergel et al. 2003;2007), large-scale structure
(LSS) (Tegmark et al. 2004; Eisenstein et al. 2005) and the
distance-redshift relation to X-ray galaxy clusters (Allen et al.
2004; 2007) also confirm that the universe is accelerating. Possible
explanations for the acceleration have been proposed. A negative
pressure term called dark energy is taken into account, such as the
cosmological constant model with equation of state $w=p/\rho =-1$
(Weinberg 1989), an evolving scalar field (Peeble \& Ratra, 1988,
Caldwell et al. 1998), the phantom energy for which the sum of the
pressure and energy density is negative, and the Chaplygin gas
(Kamenshchik et al. 2001). All the above models for acceleration are
obtained by introducing a new energy component called dark energy.
Alternative models, in which gravity is modified, can also drive the
universe acceleration, e.g., the Dvali-Gabadadze-Porrati (DGP) model
(Dvali et al. 2000; Deffayet et al. 2002), Cardassian expansion
model (Freese \& Lewis 2002; Wang et al. 2003), and the f(R) gravity
model (Vollick 2003; Carroll et al. 2004).

These two families of models, dark energy and modified gravity, are
fundamentally different. An important question is whether it is
possible to distinguish between the modified gravity and dark energy
models that have nearly the same cosmic expansion history. Many
works have been done on this topic. A usually-discussed quantity is
the growth rate of cosmological density perturbations, which should
be different in the models depending on different gravity theory
even if they have an identical cosmic expansion history. Recently,
there have been extensive discussions on discriminating dark energy
and modified gravity models using the matter density perturbations
growth factor (Linder 2005). But Kunz and Sapone (2007) demonstrated
that the growth factor is not sufficient to distinguish between
modified gravity and dark energy (Kunz \& Sapone 2007). They found
that a generalized dark energy model can match the growth rate of
the Dvali-Gabadadze-Porrati model and reproduce the 3+1 dimensional
metric perturbations.

On the other hand, the statefinder pair ($r,s$) has also been
proposed to distinguish between the models, where $r\equiv
\dot{\ddot{a}}/aH^3$ and $s\equiv (r-1)/3(q-1/2)$. Sahni et al.
(2003) demonstrated that the statefinder diagnostic could
effectively discriminate different forms of dark energy (Sahni et
al. 2003). Alam et al. (2003) investigated the cosmological
constant, quintessence, Chaplygin gas, and braneworld models using
the statefinder diagnostic, and found that the statefinder pair
could differentiate these models (Alam et al. 2003). Different
cosmological models exhibit qualitatively different trajectories of
evolution in the $r-s$ plane. The statefinder diagnostic has been
extensively used in many models (Gorini et al. 2003). But the
statefinder pair is difficult to measure by cosmological
observations (Visser 2004; Catto\"{e}n \& Visser 2007). The present
values of cosmographic parameters can be determined from
observations (Riess et al. 2004; Visser 2004). Caldwell \&
Kamionkowski (2004) showed the jerk parameter could probe the
spatial curvature of the universe (Caldwell \& Kamionkowski 2004).
The deceleration, jerk and snap parameters are related to the
second, third and fourth derivative of the scale factor
respectively. Visser (2004) expanded the Hubble law to fourth order
in redshift including the snap parameter and put constraints on the
deceleration and jerk parameters using SNe Ia (Visser 2004). Rapetti
et al. (2007) constrained the deceleration and jerk parameters from
SNe Ia and X-ray cluster gas mass fraction measurements. For a
redshift range of SNe Ia the terms beyond the cubic power of Hubble
law can be neglected. In order to put a narrow constraint on the
snap parameter, we need high-redshift objects. GRBs may be a useful
tool. GRBs can be detectable out to very high redshifts (Ciardi \&
Loeb 2000). The farthest burst detected so far is GRB 090423, which
is at $z=8.2$ (Olivares et al. 2009). A lot of work in this
so-called {\em GRB cosmology} has been published (Dai, Liang \& Xu
2004; Ghirlanda et al. 2004; Di Girolamo et al. 2005; Firmani et al.
2005; Friedman \& Bloom 2005; Lamb et al. 2005; Liang \& Zhang 2005,
2006; Xu, Dai \& Liang 2005; Wang \& Dai 2006; Schaefer 2007; Wright
2007; Wang, Dai \& Zhu 2007; Gong \& Chen 2007; Li et al. 2008;
Liang et al. 2008; Qi, Wang \& Lu 2008a,b; Basilakos \&
Perivolaropoulos 2008; Kodama et al. 2008; Wang, Dai \& Qi 2009).
Very recently, Schaefer (2007) used 69 GRBs and five relations to
build the Hubble diagram out to $z=6.60$ and discussed the
properties of dark energy in several dark energy models (Schaefer
2007). He found that the GRB Hubble diagram is consistent with the
concordance cosmology. Liang et al.(2008) calibrated the luminosity
relations of GRBs by interpolating from the Hubble diagram of SNe Ia
at $z<1.4$ with the assumption that objects at the same redshift
should have the same luminosity distance (Liang et al. 2008). This
method is model-independent. More recently, Capozziello \& Izzo
(2008) used the Liang et al. (2008) results to constrain the
cosmographic parameters and found the results calibrated by SNe Ia
data, agree with the $\Lambda$CDM model. Cardone et al. (2009) used
83 GRBs and six correlations to build the Hubble diagram. Butler et
la. (2009) found a real, intrinsic correlation between $E_{\rm iso}$
and $E_{\rm peak}$ using latest Swift GRB sample.

Riess et al. (2004) found that the jerk $j_0$ is positive at the
92\% confidence level based on their ``gold" dataset and is positive
at the 95\% confidence level based on their ``gold+silver" dataset.
Neither explicit upper bounds are given for the jerk nor are any
constraints placed on the snap $s_0$. Rapetti et al. (2007) measured
$q_0=-0.81\pm 0.14$ and $j=2.16^{+0.81}_{-0.75}$ in a flat model
with constant jerk (Rapetti et al. 2007). Capozziello \& Izzo (2008)
used 27 GRBs to derive the values of the cosmographic parameters.
They found $q_0=-0.78\pm 0.20$, $j_0=0.62\pm 0.86$ and $s_0=8.32\pm
12.16$. In this paper, we use more GRB data to constrain the
cosmography parameters in several dark energy and modified gravity
models.

In this paper, we calibrate the luminosity relations of GRBs using
SNe Ia and calculate the deceleration, jerk and snap parameters of
several dark energy and modified gravity models using SNe Ia and
GRBs. We also use a model-independent method to constrain the
cosmographic parameters. We find that in some models the jerk
parameter is almost equal to each other. So this parameter is not
used to distinguish between the models. However, the snap parameter
in all the models is different, so we can distinguish between the
models using the snap parameter.

The structure of this paper is organized as follows. In section 2 we
introduce the hubble, deceleration, jerk and snap parameters. In
section 3 we derive expressions of cosmographic parameters of the
Hubble law in several dark energy models. In section 4 we present
expressions of cosmographic parameters of the Hubble law in modified
gravity models. The constraints on models parameters and
cosmographic parameters of the Hubble law are given in section 5.
Finally, section 6 contains conclusions and discussions.

\section{Hubble, deceleration, jerk and snap parameters}
The expansion rate of the Universe can be written in terms of the
Hubble parameter, $H=\dot{a}/a$, where $a$ is the scale factor and
$\dot{a}$ is its first derivative with respect to time. As we known
that $q$ is the deceleration parameter, related to the second
derivative of the scale factor, $j$ is the so-called ``jerk'' or
statefinder parameter, related to the third derivative of the scale
factor, and $s$ is the so-called ``snap'' parameter, which is
related to the fourth derivative of the scale factor. These
quantities are defined as
\begin{equation}
q=-\frac{1}{H^2}\frac{\ddot{a}}{a};
\end{equation}
\begin{equation}
j=\frac{1}{H^3}\frac{\dot{\ddot{a}}}{a};
\end{equation}
\begin{equation}
s=\frac{1}{H^4}\frac{\ddot{\ddot{a}}}{a}.
\end{equation}
The deceleration, jerk and snap parameters are dimensionless, and a
Taylor expansion of the scale factor around $t_0$ provides
\begin{eqnarray}
a(t)=a_0 \left\{1+H_0(t-t_0)-\frac{1}{2}q_0H_{0}^{2}(t-t_0)^2
+\frac{1}{3!}j_0H_{0}^{3}(t-t_0)^3\right.
\nonumber\\
+\frac{1}{4!}s_0H_{0}^{4}(t-t_0)^4+O[(t-t_0)^5]\},
\end{eqnarray}
and so the luminosity distance \beqa
d_L={{c}\over{H_0}}\left\{z+{{1\over2}(1-q_0)}z^2-{{1}\over{6}}
\left(1-q_0-3q_0^2+j_0\right)z^3\right.
\nonumber\\
+{{1}\over{24}}\left[2-2q_0-15q_{0}^{2}-15q_0^3+5j_0 +10q_0
j_0+s_0\right]z^4 +O(z^5)\}, \eeqa (Visser 2004). For the redshift
range of SNe Ia the terms beyond the cubic power in Eq. (5) can be
neglected. If models have the same deceleration and jerk parameters,
we can see degeneracy of these models from Eq. (5). Therefore we
must measure the snap parameters to distinguish between the models.
This needs high-redshift objects. The relations among the
$q(z),j(z)$ and $s(z)$ are \be j(z)=q(z)+2q^2(z)+(1+z){dq\over
dz}(z); \ee \be s(z)=-(1+z){dj\over dz}(z)-2j(z)-3j(z)q(z). \ee The
Friedmann equation is
\begin{equation}
H^2  = (\frac{{\dot a}}{a})^2  = \frac{{8\pi G}}{3}\sum\limits_i
{\rho _i }.
\end{equation}
From Einstein's equations, we can obtain the dynamical equation of
universe
\begin{equation}
 \frac{{\ddot a}}{a} =  - \frac{{4\pi G}}{3}\sum\limits_i {(\rho _i
+ 3P_i )}.\label{einstein}
\end{equation}
The conservation equation is \be \dot \rho _i  + 3H(\rho _i  + P_i
)= 0. \ee In order to derive the jerk and snap parameter, we
differentiate Eq.(9)
\begin{eqnarray}
\dot{\ddot{a}}& =& - \frac{{4\pi G}}{3}\sum\limits_i {[\dot a\rho _i (1 + 3w_i ) + a\dot \rho _i (1 + 3w_i ) + a\rho _i  \times 3\dot w_i ]} \nonumber  \\
              & =& - \frac{{4\pi G}}{3}\sum\limits_i [aH\rho _i (1 +3w_i ) - 3Ha\rho _i
(1 + w_i )(1 + 3w_i ) \nonumber \\
& & + a \rho_i  \times 3\dot{w_i}],
\end{eqnarray}

\begin{eqnarray}
 \ddot{\ddot a}& =& - \frac{{4\pi G}}{3}\sum\limits_i {\frac{{d[\dot a\rho _i (1 + 3w_i ) + a\dot \rho _i (1 + 3w_i ) + a\rho _i  \times 3\dot w_i ]}}{{dt}}}  \nonumber \\
  &=&  - \frac{{4\pi G}}{3}\sum\limits_i [\ddot a \rho _i (1 + 3w_i ) + 2\dot a \dot \rho_i  (1 + 3w_i ) + 6\dot a \rho _i \dot w_i  \nonumber \\
 & &  + a \ddot \rho_i  (1 + 3w_i )+ 6a \dot \rho_i \dot w_i  + a\rho _i  \times 3\ddot w_i ].
 \end{eqnarray}

\section{Dark energy models}
\subsection{$w(z)$ parameterization model}
We first consider the dark energy with a constant equation of state.
\begin{equation}
w(z)=w_0
\end{equation}
For this model, we obtain \be q_0^{\rm XCDM}={3 \over
2}\left[1+w_0(1- \Omega_M)\right]-1, \ee \be \left.{d q\over d
z}\right|_0^{\rm XCDM}={9\over2} w_0^2 (1-\Omega_M) \Omega_M, \ee
\be \left.{d j\over d z}\right|_0^{\rm XCDM}= -{27 \over
2}w_0^2(1+w_0) (\Omega_M-1)\Omega_M, \ee

\be j_0^{\rm
XCDM}=\frac{1}{2}(2+9(1-\Omega_M)w_0+9(1-\Omega_M)w_0^2), \ee
\begin{eqnarray}
s_0^{\rm XCDM} &=&
\frac{1}{4}(-14-81(1-\Omega_M)w_0-9(16-19\Omega_M+3\Omega_M^2)w_0^2
\nonumber \\
& & -27(3-4\Omega_M+\Omega_M^2)w_0^3).
\end{eqnarray}
These expressions are consistent with Bertolami \& Silva (2006).

A more interesting approach to explore dark energy is to use
time-dependent dark energy model. The simplest parameterization
including two parameters is (Maor et al. 2001; Weller \& Albrecht
2001)
\begin{equation}
w(z)=w_{0}+w_{1}z.
\end{equation}
In this dark energy model the luminosity distance is (Linder 2003)
\begin{eqnarray}
d_{L}&=& cH_{0}^{-1}(1+z)\int_{0}^{z}dz[(1+z)^{3}\Omega_{M}
\nonumber \\
& & +(1-\Omega_{M})(1+z)^{3(1+w_{0}-w_{1})}e^{3w_{1}z}]^{-1/2}.
\end{eqnarray}
\be q_0^{\rm WCDM}=\frac{1}{2}+\frac{3}{2}(1-\Omega_M)w_0 \ee \be
j_0^{\rm
WCDM}=\frac{1}{2}(2+9(1-\Omega_M)w_0+9(1-\Omega_M)w_0^2+3(1-\Omega_M)w_1)
\ee
\begin{eqnarray}
s_0^{\rm WCDM} &=& \frac{1}{4}(-14-9(16-19\Omega_M+3\Omega_M^2)w_0^2
\nonumber \\
& & -27(3-4\Omega_M+\Omega_M^2)w_0^3-45(1-\Omega_M)w_1 \nonumber \\
& & +9w_0(1-\Omega_M)(-9-(\Omega_M-7)w_1))
\end{eqnarray}

We consider the Chevallier-Polarski-Linder parameterization
(Chevallier \& Polarski 2001; Linder 2003)
\begin{equation}
 w(z)=w_{0}+\frac{w_{1}z}{1+z}.
\end{equation}
The luminosity distance is (Chevallier \& Polarski 2001; Linder
2003)
\begin{eqnarray}
d_{L}&=& cH_{0}^{-1}(1+z)\int_{0}^{z}dz[(1+z)^{3}\Omega_{M}
\nonumber \\
& &
+(1-\Omega_{M})(1+z)^{3(1+w_{0}+w_{1})}e^{-3w_{1}z/(1+z)}]^{-1/2}.
\end{eqnarray}
The cosmographic parameters are: \be q_0^{\rm
CPL}=\frac{1}{2}+\frac{3}{2}(1-\Omega_M)w_0, \ee \be j_0^{\rm
CPL}=\frac{1}{2}(2+9(1-\Omega_M)w_0+9(1-\Omega_M)w_0^2+3(1-\Omega_M)w_1),
\ee
\begin{eqnarray}
s_0^{\rm CPL} &=& \frac{1}{4}(-14-9(16-19\Omega_M+3\Omega_M^2)w_0^2
\nonumber \\
& & -27(3-4\Omega_M+\Omega_M^2)w_0^3-33(1-\Omega_M)w_1 \nonumber \\
& & +9w_0(1-\Omega_M)(-9-(\Omega_M-7)w_1)).
\end{eqnarray}
Capozziello, Cardone \& Salzano (2008) and Capozziello \& Izzo
(2008) also derived cosmographic parameters in this model. Our
results are consistent with theirs.

\subsection{generalized Chaplygin gas model}
We consider the generalized Chaplygin gas (GCG) model, which is
characterized by the equation of state
\begin{equation}
p_{\rm GCG}=-A/\rho_{\rm GCG}^{\alpha}.
\end{equation}
We can integrate the conservation equation for generalized Chaplygin
gas, leading to
\begin{equation}
\rho_{\rm GCG}=\rho_{\rm
GCG0}[A_s+(1-A_s)a^{-3(1+\alpha)}]^{1/(1+\alpha)} \label{GCG}
\end{equation}
where $\rho_{\rm Ch0}$ is the energy density of GCG today, and
$A_s=A/\rho_{\rm Ch0}^{1+\alpha}$. The attractive feature of the
model is that it can unify dark energy and dark matter. The reason
is that, from Eq. (30), the GCG behaves as dustlike matter at an
early epoch and as a cosmological constant at a later epoch
(Kamenshchik et al. 2001; Bento et al. 2002). The Friedmann equation
can be expressed as
\begin{equation}
H^2(z,H_0,A_s,\alpha)=H_0^2E^2(z,A_s,\alpha),
\end{equation}
where
\begin{equation}
E^2(z,A_s,\alpha)=\Omega_b(1+z)^3+(1-\Omega_b)[A_s+(1-A_s)(1+z)^{3(1+\alpha)}]^{\frac{1}{1+\alpha}},
\end{equation}
$\Omega_b$ is the density parameter of the baryonic matter. The
luminosity distance is
\begin{eqnarray}
d_{L}&=& cH_{0}^{-1}(1+z)\int_{0}^{z}dz
\{(1+z)^{3}\Omega_{b}+(1-\Omega_{b}) \nonumber \\
& & [A_s+(1-A_s)(1+z)^{3(1+\alpha)}]^{\frac{1}{1+\alpha}}\}^{-1/2}.
\end{eqnarray}
For the GCG model we obtain (Bertolami \& Silva 2006; Wang, Dai \&
Qi 2009) \be q_0^{\rm GCG}={3\over2} (1-A_s)-1 \ee \be \left.{d
q\over d z}\right|_0^{\rm GCG}={9\over2}A_s(1-A_s)(1+\alpha), \ee
\be \left.{d j\over d z}\right|_0^{\rm GCG}= -{27 \over
2}\alpha(1+\alpha)(2A_s-1) \left(A_s-1\right)A_s, \ee \be j_0^{\rm
GCG}=\frac{3}{4}(1-A_s)(1+(3+6 \alpha)A_s), \ee \be s_0^{\rm
GCG}=\frac{3}{8}(A_s-1)(7+6(2+\alpha-6\alpha^2)A_s+9(-3+2\alpha+8\alpha^2)A_s^2).
\ee

\section{Modified gravity models}
\subsection{Cardassian expansion model}
The original Cardassian model was first introduced in (Freese \&
Lewis 2002) as a possible alternative to explain the acceleration
of the universe. They modified the Friedmann equation as
\begin{eqnarray} \label{Carda}
H^2=\frac{8\pi G}{3}\rho_m+B\rho_m^n.
\end{eqnarray}
This model has no energy component besides ordinary matter. If we
consider a spatially flat FRW universe, the Friedmann equation is
modified as Eq. (39). The universe undergoes acceleration requires
$n < 2/3$. If $n=0$, it is the same as the cosmological constant
universe. We can obtain $H(z)$ by using Eq. (\ref{Carda}) and
$\rho_m=\rho_{m}(1+z)^3=\Omega_{m}\rho_c(1+z)^3$,
\begin{equation} \label{CardaH}
H(z)^2=H_0^2[\Omega_{m}(1+z)^3+(1-\Omega_{m})(1+z)^{3n}],
\end{equation}
where $\rho_c=3H_0^2/8\pi G$ is the critical density of the
universe. The luminosity distance in this model is
\begin{equation}
d_{L}=cH_{0}^{-1}(1+z)\int_{0}^{z}dz[(1+z)^{3}\Omega_{m}+(1-\Omega_{m})(1+z)^{3n}]^{-1/2}.
\end{equation}
For the Cardassian expansion model, we obtain \be q_0^{\rm
Card}={1 \over 2}+{3 \over 2}(1-n)(\Omega_M-1), \ee \be \left.{d
q\over d z}\right|_0^{\rm Card}={9\over2} (n-1)^2 (1-\Omega_M)
\Omega_M, \ee \be \left.{d j\over d z}\right|_0^{\rm Card}= {27
\over 2}(n-1)^2 (1-\Omega_M)\Omega_M n, \ee

\be j_0^{\rm Card}=\frac{1}{2}(2+9n(\Omega_M-1)+9n^2(1-\Omega_M))
\ee

\begin{eqnarray}
s_0^{\rm Card}&=&
\frac{1}{4}(4-18\Omega_M-27n^3(3-4\Omega_M+\Omega_M^2)
\nonumber \\
& & -9n(4-7\Omega_M+3\Omega_M^2)+9n^2(11-17\Omega_M+6\Omega_M^2)).
\end{eqnarray}

\subsection{Dvali-Gabadadze-Porrati model}
In the DGP model the modified Friedmann equation due to the
presence of an infinite-volume extra dimension is (Deffayet et al.
2002)
\begin{equation}
\label{eq:ansatz} H^2 = H_0^2 \left[
    \Omega_k(1+z)^2+\left(\sqrt{\Omega_{r_c}}+
    \sqrt{\Omega_{r_c}+\Omega_m (1+z)^3}\right)^2
        \right],
\end{equation}
where the bulk-induced term, $\Omega_{r_c}$, is defined as
\begin{equation}
\label{eq:omegarc} \Omega_{r_c} \equiv 1/4r_c^2H_0^2.
\end{equation}
For a flat universe, $\Omega_k=0$. In the above equation, $r_c$ is
the crossover scale beyond which the gravitational force follows the
5-dimensional $1/r^3$ behavior. Note that on short length scales $r
\ll r_c$ (at early times) the gravitational force follows the usual
four-dimensional $1/r^2$ behavior. For a spatially flat universe,
$\Omega_{r_c}=(1-\Omega_m)^2/4$. We obtain \be q_0^{\rm DGP}={1
\over 2}+{3 \over 2}\frac{\Omega_M-1}{1+\Omega_M}, \ee \be \left.{d
q\over d z}\right|_0^{\rm DGP}=\frac{9\Omega_M
(1-\Omega_M)}{(1+\Omega_M)^3}, \ee \be \left.{d j\over d
z}\right|_0^{\rm DGP}= \frac{54\Omega_M^3
(1-\Omega_M)}{(1+\Omega_M)^5}, \ee

\be j_0^{\rm
DGP}=\frac{1+3\Omega_M-6\Omega_M^2+10\Omega_M^3}{(1+\Omega_M)^3},
\ee

\be s_0^{\rm
DGP}=\frac{1-4\Omega_M-35\Omega_M^2-26\Omega_M^3+32\Omega_M^4-80\Omega_M^5}{(1+\Omega_M)^5}.
\ee

\subsection{$f(R)$ gravity}
$f(R)$ gravity models, in which the gravitational Lagrangian is a
function of the curvature scalar $R$, also can explain current
cosmic acceleration (Vollick 2003; Carroll et al. 2004; Capozziello
et al. 2009). Poplawski (2006) derived a quite complicated
expression of jerk parameter in $f(R)=R-\frac{\alpha^2}{3R}$
(Poplawski 2006):
\begin{eqnarray}
& & j'=[(\phi f'-3f)(2f'^2+\phi f'f''-6ff'')(30\phi^3 f'^2 f''^2+10\phi^3 f'^3 f''' \nonumber \\
& & -150\phi^2 ff'f''^2-37\phi^2 f'^3 f''-75\phi^2 ff'^2 f'''-8\phi f'^4+24ff'^3 \nonumber \\
& & +189\phi f^2 f''^2+189\phi f^2 f'f'''+192\phi ff'^2 f''-162f^3 f''' \nonumber \\
& & -267f^2 f'f'')-(2\phi^2 f'^4+10\phi^3 f'^3 f''-75\phi^2 f'^2 ff''-12\phi ff'^3 \nonumber \\
& & +189\phi f^2 f'f''-162f^3 f'')\times(3\phi^2 f'f''^2-15\phi ff''^2-8f'^3 \nonumber \\
& & +18f^2 f'^2+27ff'f''-\phi f'^2f''+\phi^2 f'^2 f'''-9\phi ff'f''' \nonumber \\
& & +18f^2f''')]\times[(\phi f'-3f)^3(2f'^2+\phi
f'f''-6ff'')^2]^{-1}. \label{sn3}
\end{eqnarray}
The snap parameter in this model is (Poplawski 2007)
\begin{equation}
s=j'\frac{6f'(\phi f'-2f)}{(2f'^2+\phi f'f''-6ff'')}-j\frac{8\phi
f'-15f}{\phi f'-3f}. \label{sn4}
\end{equation}
Poplawski (2007) calculated $q_0=-0.67^{+0.06}_{-0.03}$,
$j_0=1.01^{+0.08}_{-0.21}$ and $s_0=-0.22^{+0.21}_{-0.19}$. These
expressions of the jerk and snap parameters are only valid in
Palatini variational principle. A generic formulae of cosmographic
parameter are derived by Capozziello, Cardone \& Salzano (2008) (for
more details, see Equations (23)-(33) in their paper). They also
gave the best fitted value: $q_0=-0.55\pm0.38$, $j_0=1.0\pm5.4$ and
$s_0=-0.35\pm28.1$ using SNe Ia. In this paper, we only use
Poplawski (2006) as an example for the $f(R)$ gravity.

\section{Constraints from SNe Ia and GRBs}\label{analysis}
Davis et al. (2007) fitted the SNe Ia dataset that include 60
ESSENCE SNe Ia (WoodVasey et al. 2007), 57 SNe Ia from Super-Nova
Legacy Survey (SNLS) (Astier et al. 2006), 45 nearby SNe Ia and 30
SNe Ia detected by HST (Riess et al. 2007) with the MCLS2K2 method.
With the luminosity distance $d_{L}$ in units of megaparsecs, the
predicted distance modulus is
\begin{equation}
\mu=5\log(d_{L})+25.
\end{equation}
The likelihood functions can be determined from $\chi^{2}$
statistic,
\begin{equation}
\chi^{2}_{\rm SNe}=\sum_{i=1}^{N}
\frac{[\mu_{i}(z_{i})-\mu_{0,i}]^{2}}{\sigma_{\mu_{0,i}}^{2}+\sigma_{\nu}^{2}},
\end{equation}
where $\sigma_{\nu}$ is the dispersion in the supernova redshift
(transformed to distance modulus) due to peculiar velocities,
$\mu_{0,i}$ is the observational distance modulus, and
$\sigma_{\mu_{0,i}}$ is the uncertainty in the individual distance
moduli. The confidence regions can be found through marginalizing
the likelihood functions over $H_{0}$ (i.e., integrating the
probability density $p\propto\exp^{-\chi^{2}/2}$ for all values of
$H_{0}$).

We use the calibration results obtained by using the interpolation
methods directly from SNe Ia data (Liang et al. 2008). The
calibrated luminosity relations are completely cosmology
independent. We assume these relations do not evolve with redshift
and are valid in $z>1.40$. The luminosity or energy of GRB can be
calculated. So the luminosity distances and distance modulus can
be obtained. After obtaining the distance modulus of each burst
using one of these relations, we use the same method as Schaefer
(2007) to calculate the real distance modulus,
\begin{equation}
\mu_{\rm fit}=(\sum_i \mu_i/\sigma_{\mu_i}^2)/(\sum_i
\sigma_{\mu_i}^{-2}),
\end{equation}
where the summation runs from $1-5$ over the relations with
available data, $\mu_i$ is the best estimated distance modulus from
the $i$-th relation, and $\sigma_{\mu_i}$ is the corresponding
uncertainty. The uncertainty of the distance modulus for each burst
is
\begin{equation}
\sigma_{\mu_{\rm fit}}=(\sum_i \sigma_{\mu_i}^{-2})^{-1/2}.
\end{equation} The $\chi^2$ value
is
\begin{equation}
\chi^{2}_{\rm GRB}=\sum_{i=1}^{N} \frac{[\mu_{i}(z_{i})-\mu_{{
fit},i}]^{2}}{\sigma_{\mu_{{\rm fit},i}}^{2}},
\end{equation}
where $\mu_{{\rm fit},i}$ and $\sigma_{\mu_{{\rm fit},i}}$ are the
fitted distance modulus and its error.

We combine SNe Ia and GRBs by multiplying the likelihood functions.
The total $\chi^2$ value is $\chi^2_{\rm total}=\chi^2_{\rm
SNe}+\chi^2_{\rm GRB}$. The best fitted value is obtained by
minimizing $\chi^2_{\rm total}$.

\subsection{Constraints on cosmographic parameters} In our
analysis, we consider the flat cosmology. We use $h=0.72\pm0.08$
from the {\em Hubble Space Telescope key projects} (Freedman et al.
2001). Riess et al. (2009) used the old distance ladder and observed
Cepheids in the near-infrared  where they are less sensitive to dust
and found $h=0.742\pm0.036$.

Let us first consider observational constraints on dark energy
models. In Fig.1, we show the distribution probability a function of
$\Omega_M$ in the flat $\Lambda$CDM model from SNe Ia and GRBs. From
this figure, we have $\Omega_M=0.27\pm0.04$. The cosmographic
parameters in $\Lambda$CDM model are $q_0 = -1 +
\frac{3}{2}\Omega_M$, $j_0=1.0$ and $s_0 = 1 - \frac{9}{2}\Omega_M$.
We can obtain $q_0=-0.60\pm0.06$, $j_0=1.0$ and $s_0=-0.22\pm0.18$.

\begin{figure}[htbp]
\includegraphics[width=9cm]{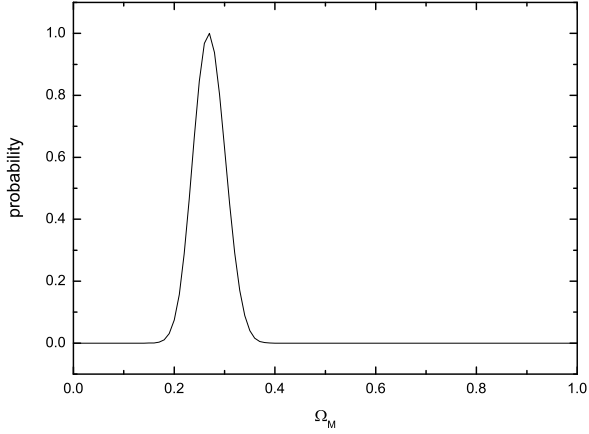}
\caption{Luminosity distance - redshift diagram.  The circles are
the GRBs. The solid line is the result of our fitting}
\label{fig:no1}
\end{figure}

In Fig.2 we present constraints on $\Omega_{M}$ and $w$ from
$1\sigma$ to $3\sigma$ using 192 SNe Ia and 69 GRBs in the $w=w_0$
model. We measure $\Omega_M=0.29_{-0.14}^{+0.11}$ and
$w_0=-1.04_{-0.52}^{+0.32}$. The cosmographic parameters in the
$w=w_0$ model are $q_0=-0.61^{+0.38}_{-0.60}$,
$j_0=1.13^{+1.10}_{-1.79}$ and $s_0=-0.08^{+1.75}_{-2.75}$.

\begin{figure}[htbp]
\includegraphics[width=9cm]{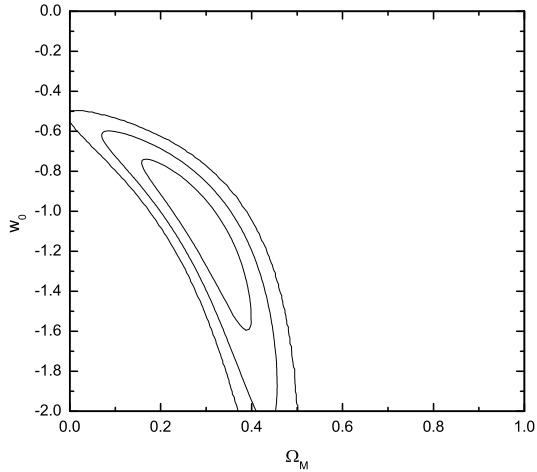}
 \caption{\label{Fig2} Constraints on $\Omega_{M}$ and $w$ from $1\sigma$ to $3\sigma$ using 192 SNe
 Ia in the $w=w_0$ model.
 }
\end{figure}

In Fig.3 we present constraints on $w_0$ and $w_1$ from $1\sigma$ to
$3\sigma$ using 192 SNe Ia and 69 GRBs in the $w=w_0+w_1 z$ model.
The values of parameters are $w_0=-1.14\pm0.19$ and
$w_1=0.63\pm0.44$. The cosmographic parameters are
$q_0=-0.75\pm0.21$, $j_0=2.21\pm0.93$ and $s_0=-12.25\pm9.18$.

\begin{figure}[htbp]
\includegraphics[width=9cm]{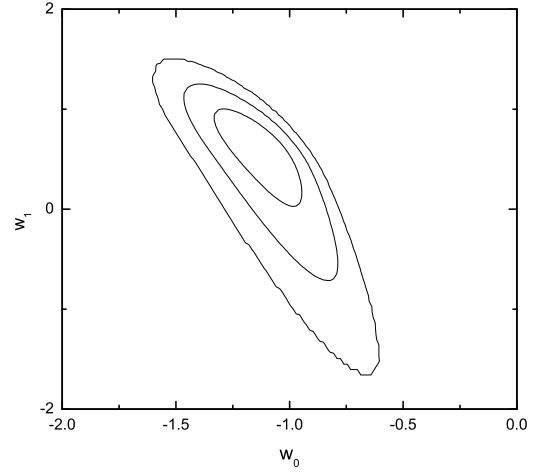}
 \caption{\label{Fig3} The same as Fig.2 but in the $w=w_0+w_1 z$
 model.
 }
\end{figure}

In Fig.4 we present constraints on $w_0$ and $w_1$ from $1\sigma$ to
$3\sigma$ using 192 SNe Ia and 69 GRBs in the $w=w_0+w_1 z/(1+z)$
model. We measure $w_0=-1.22\pm0.30$ and $w_1=1.6^{+1.20}_{-1.10}$.
The cosmographic parameters are $q_0=-0.90\pm0.33$,
$j_0=3.93^{+1.93}_{-2.09}$ and $s_0=-25.52^{+27.33}_{-25.33}$.

\begin{figure}[htbp]
\includegraphics[width=0.5\textwidth]{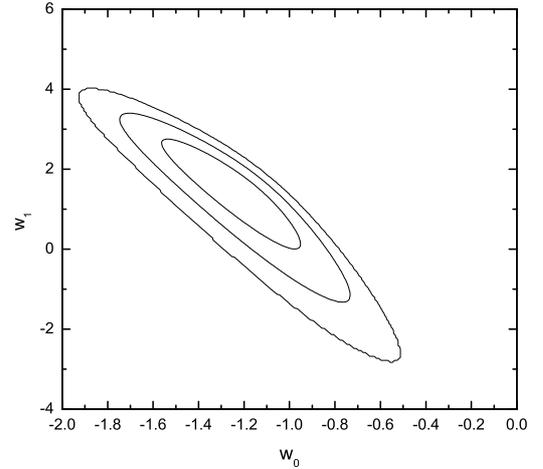}
 \caption{\label{Fig4} The same as Fig.2 but in the $w=w_0+w_1 z/(1+z)$
 model.
 }
\end{figure}

Fig.5 shows constraints on $A_s$ and $\alpha$ from $1\sigma$ to
$3\sigma$ using SNe Ia and GRBs in the GCG model. The parameters are
$A_s=0.79\pm0.12$ and $\alpha=0.25_{-0.75}^{+0.95}$. The
cosmographic parameters are $q_0=-0.695\pm0.18$,
$j_0=1.18^{+0.78}_{-0.65}$ and $s_0=-0.37_{-1.48}^{+1.85}$.

\begin{figure}[htbp]
\includegraphics[width=9cm]{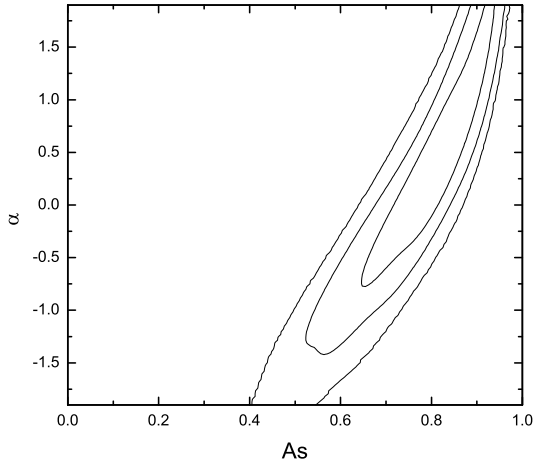}
 \caption{\label{Fig5} The same as Fig.2 but in the GCG
 model.
 }
\end{figure}

In Fig.6 we present constraints on $\Omega_M$ and $n$ from $1\sigma$
to $3\sigma$ using 192 SNe Ia and 69 GRBs in the Cardassian
expansion model. We measure $\Omega_M=0.29\pm0.11$ and
$n=-0.07^{+0.34}_{-0.46}$. The cosmographic parameters are
$q_0=-0.67^{+0.35}_{-0.40}$, $j_0=1.35^{+1.20}_{-1.45}$ and
$s_0=0.36^{+2.51}_{-2.85}$.

\begin{figure}[htbp]
\includegraphics[width=9cm]{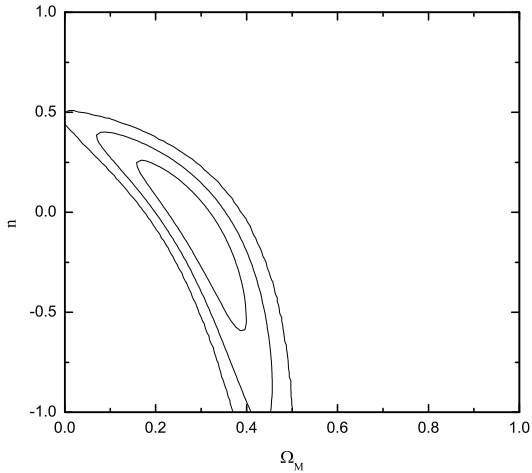}
 \caption{\label{Fig6}The same as Fig.2 but in the Cardassian expansion
 model.}
\end{figure}

Fig.7 shows constraints on $\Omega_M$ using SNe Ia and GRBs in the
DGP model. The value of $\Omega_M$ is $\Omega_M=0.20\pm0.02$. The
cosmographic parameters are $q_0=-0.50\pm0.04$, $j_0=0.83\pm0.02$
and $s_0=-0.56\pm0.12$.

\begin{figure}[htbp]
\includegraphics[width=9cm]{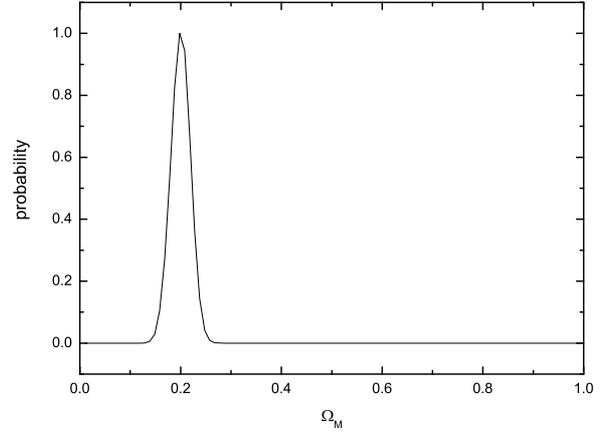}
 \caption{\label{Fig7} The same as Fig.1 but in the DGP model.
 }
\end{figure}

We directly use Eq.(5) to constrain the cosmographic parameters.
This analysis uses the FRW metric only, so we have not specified any
gravitational theory yet. The luminosity distance only depends on
redshift $z$ and cosmographic parameters. So this method is fully
model independent. We use the 192 SNe Ia and 69 GRBs and find the
best fit parameters are $q_0=-0.85 \pm 0.19$, $j_0=1.50\pm 0.80$ and
$s_0=3.49\pm 4.50$. The results are consistent with the flat
$\Lambda$CDM model.

In Table 1 we summarize the constraints on cosmographic parameters.
The deceleration and jerk parameters in the $w=w_0$, GCG, Cardassian
expansion and f(R) models are almost the same in the $1\sigma$
confidence level. These values are consistent with the deceleration
and jerk parameters of the $\Lambda$CDM model in the $1\sigma$
confidence level. So these models can not be discriminated using the
present value of the statefinder pair. However the snap parameter in
all the models is different and thus can be used to discriminate the
cosmological models. In the future, more data will give a precise
snap parameter in different models.

\section{conclusions and discussions}
The cosmic acceleration could be due to a mysterious dark energy, or
a modification of general relativity (modified gravity). In this
paper we investigate the deceleration, jerk and snap parameters in
modified gravity models and dark energy models. We calibrate the GRB
luminosity relations without assuming any cosmological models using
SNe Ia. Because gamma-ray bursts can be detected in high redshifts,
we calculate the deceleration, jerk and snap parameters using type
Ia supernovae and gamma-ray bursts. GRBs can extend the redshift -
distance relation up to high redshifts. We find that the
deceleration and jerk parameters in the $w=w_0$, GCG, Cardassian
expansion and f(R) models are almost the same in the $1\sigma$
confidence level. So these models can not be discriminated using the
present value of the statefinder pair. We find that the dark energy
models and modified gravity models could be distinguished between by
the snap parameter. Using the model-independent constraints on
cosmographic parameters, we find the $\Lambda$CDM model is
consistent with the current data.

\section*{Acknowledgements}
This work is supported by the National Natural Science Foundation of
China (grants 10233010, 10221001 and 10873009) and the National
Basic Research Program of China (973 program) No. 2007CB815404. F.
Y. Wang was also supported by the Jiangsu Project Innovation for PhD
Candidates (CX07B-039z).

\clearpage

\begin{table}
\caption{The cosmographic parameters value}
\begin{tabular}{cccc}
\hline\hline model & $q_0$ & $j_0$ & $s_0$ \\
\hline
$\Lambda$CDM &$-0.60\pm0.06$&$1.0\pm0.11$&$-0.22\pm0.18$ \\
$w=w_0$ & $-0.61^{+0.38}_{-0.60}$ & $1.13^{+1.10}_{-1.79}$ &
$-0.08^{+1.75}_{-2.75}$ \\
$w=w_0+w_1 z$ & $-0.75\pm0.21$ & $2.21\pm0.93$ & $-12.25\pm9.18$ \\
$w=w_0+\frac{w_1 z}{1+z}$ & $-0.90\pm0.33$ & $3.93^{+1.93}_{-2.90}$
& $-25.52^{+27.33}_{-25.33}$ \\
GCG & $-0.70\pm0.18$ & $1.18^{+0.78}_{-0.65}$ &
$-0.37^{+1.35}_{-1.48}$ \\
Cardassian & $-0.67^{+0.35}_{-0.40}$ & $1.20^{+1.20}_{-1.45}$ &
$0.22^{+2.51}_{-2.85}$ \\
DGP & $-0.50\pm0.04$ & $0.83\pm0.02$ & $-0.56\pm0.12$ \\
f(R) & $-0.67^{+0.06}_{-0.03}$ & $1.01^{+0.08}_{-0.21}$ & $-0.22^{+0.21}_{-0.19}$ \\
 \hline
\end{tabular}

\end{table}


\begin{thebibliography}{}
\bibitem[\protect\citeauthoryear{}{2003}]{} Alam, U., et al. 2003, MNRAS, 344, 1057
\bibitem[\protect\citeauthoryear{}{2004}]{} Allen, S. W., et al. 2004, MNRAS, 353, 457
\bibitem[\protect\citeauthoryear{}{2008}]{} Allen, S. W., et al. 2008, MNRAS, 383, 879
\bibitem[\protect\citeauthoryear{}{2006}]{} Astier, P. et al. 2006, A\&A, 447, 31
\bibitem[\protect\citeauthoryear{}{2006}]{} Bertolami, O. \& Silva, P. T. 2006, MNRAS, 356,
1149
\bibitem[\protect\citeauthoryear{}{2008}]{} Basilakos, S. \& Perivolaropoulos, L. arXiv:
0805.0875
\bibitem[\protect\citeauthoryear{}{2003}]{} Bennett, C. L. et al. 2003, ApJS, 148, 97
\bibitem[\protect\citeauthoryear{}{2002}]{} Bento, M. C., Bertolami, O. \& Sen, A. A. 2002, Phys. Rev. D, 66, 043507
\bibitem[\protect\citeauthoryear{}{2009}]{} Butler, N. R., Bloom, J. S. \& Poznanski, D., 2009,
arXiv: 0910.3341
\bibitem[\protect\citeauthoryear{}{1998}]{} Caldwell, R. R., Dave, R. \& Steinhardt, P. J. 1998, Phys. Rev. Lett, 80, 1582
\bibitem[\protect\citeauthoryear{}{2004}]{} Caldwell, R. R., \& Kamionkowski, M., 2004, JCAP, 0409, 009
\bibitem[\protect\citeauthoryear{}{2008}]{} Capozziello, S., Cardone, V. F. \& Salzano, V.
2008, Phys. Rev. D, 78, 063504
\bibitem[\protect\citeauthoryear{}{2008}]{} Capozziello, S. \& Izzo, L., 2008, A\&A, 490, 31
\bibitem[\protect\citeauthoryear{}{2009}]{} Capozziello, S. et al. 2009, Phys.Lett.B., 671,
193
\bibitem[\protect\citeauthoryear{}{2009}]{} Cardone, V. F., Capozziello, S. \& Dainotti, M. G., 2009, arXiv:0901.3194
\bibitem[\protect\citeauthoryear{}{2004}]{} Carroll, S. M., et al., 2004, Phys. Rev. D, 70, 043528
\bibitem[\protect\citeauthoryear{}{2007}]{} Catto\"{e}n, C \& Visser, M., gr-qc/0703122v3
\bibitem[\protect\citeauthoryear{}{2001}]{} Chevallier, M. \& Polarski, D., 2001, Int. J. Mod. Phys. D, 10, 213
\bibitem[\protect\citeauthoryear{}{2004}]{} Dai, Z. G., Liang, E. W. \& Xu, D. 2004, ApJ, 612, L101
\bibitem[\protect\citeauthoryear{}{2007}]{} Davis, T. M., et al. 2007, ApJ, 666, 716
\bibitem[\protect\citeauthoryear{}{2002}]{} Deffayet, C, Dvali, G. R. \& Gabadadze, G. 2002, Phys. Rev. D, 65, 044023
\bibitem[\protect\citeauthoryear{}{2000}]{} Ciardi, B. \& Loeb, A., 2000, ApJ, 540, 687
\bibitem[\protect\citeauthoryear{}{2000}]{} Dvali, G. R., Gabadadze, G. \& Porrati, M., 2000, Phys. Lett. B, 485, 208
\bibitem[\protect\citeauthoryear{}{2005}]{} Di Girolamo, T. et al. 2005, JCAP, 4, 008
\bibitem[\protect\citeauthoryear{}{2005}]{} Eisenstein, D. J. et al. 2005, ApJ, 633, 560
\bibitem[\protect\citeauthoryear{}{2005}]{} Firmani, C., Ghisellini, G., Ghirlanda, G., \& Avila-Reese, V. 2005, MNRAS, 360, L1
\bibitem[\protect\citeauthoryear{}{2000}]{} Fenimore, E. E. \& Ramirez-Ruiz, E. 2000, astro-ph/0004176
\bibitem[\protect\citeauthoryear{}{2001}]{} Freedman, W. et al. 2001, ApJ. 553, 47
\bibitem[\protect\citeauthoryear{}{2002}]{} Freese, K. \& Lewis, M., 2002, Phys. Lett. B. 540, 1
\bibitem[\protect\citeauthoryear{}{2005}]{} Friedmann, A. S. \& Bloom, J. S. 2005, ApJ, 627, 1
\bibitem[\protect\citeauthoryear{}{2004}]{} Ghirlanda, G. et al. 2004, ApJ, 613, L13
\bibitem[\protect\citeauthoryear{}{2003}]{} Gorini, V., Kamenshchik, A. \& Moschella, U., 2003, Phys. Rev. D., 67, 063509
\bibitem[\protect\citeauthoryear{}{2007}]{} Huterer, D. \& Linder, E. V., 2007, Phys. Rev. D., 75, 023519.
\bibitem[\protect\citeauthoryear{}{2006}]{} Ishak, M., Upadhye, A. \& Spergel,D. N., 2006, Phys. Rev. D 74, 043513.
\bibitem[\protect\citeauthoryear{}{2001}]{} Kamenshchik, A., Moschella, U. \& Pasquier, V. 2001, Phys. Lett.B, 511, 265
\bibitem[\protect\citeauthoryear{}{2008}]{} Kodama, et al. arXiv: 0802.3428v2
\bibitem[\protect\citeauthoryear{}{2007}]{} Kunz, M. \& Sapone, D., 2007, Phys. Rev. Lett. 98, 121301
\bibitem[\protect\citeauthoryear{}{2005}]{} Lamb, D. Q. et al. 2005, astro-ph/0507362
\bibitem[\protect\citeauthoryear{}{2008}]{} Li, H., et al. 2008, Phys. Lett.B, 658, 95
\bibitem[\protect\citeauthoryear{}{2005}]{} Liang, E. W. \& Zhang, B. 2005, ApJ, 633, 611
\bibitem[\protect\citeauthoryear{}{2006}]{} Liang, E. W. \& Zhang, B. 2006, MNRAS, 369, L37
\bibitem[\protect\citeauthoryear{}{2008}]{} Liang, N., et al. arXiv: 0802.4262v4
\bibitem[\protect\citeauthoryear{}{2003}]{} Linder, E. V., 2003, Phys. Rev. Lett, 90, 091301
\bibitem[\protect\citeauthoryear{}{2005}]{} Linder, E.V., 2005, Phys. Rev. D 72, 043529.
\bibitem[\protect\citeauthoryear{}{2001}]{} Maor, I., Brustein, R. \& Steinhardt, P. J., 2001, Phys. Rev. Lett,87, 049901
\bibitem[\protect\citeauthoryear{}{1988}]{} Peebles, P. J. E. \& Ratra, B., 1988, ApJ, 325, L17
\bibitem[\protect\citeauthoryear{}{1999}]{} Perlmutter, S. et al. 1999, ApJ, 517, 565
\bibitem[\protect\citeauthoryear{}{2007}]{} Polarski, D. \& Gannouji, R.,  arXiv:0710.1510.
\bibitem[\protect\citeauthoryear{}{2006}]{} Poplawski, N. J., 2006, Phys. Lett. B. 640, 135
\bibitem[\protect\citeauthoryear{}{2007}]{} Poplawski, N. J., 2007, Class. Quantum. Grav. 24, 3013
\bibitem[\protect\citeauthoryear{}{2008}]{} Qi, S., F. Y. Wang. \& Lu, T. 2008a, A\&A, 483, 49
\bibitem[\protect\citeauthoryear{}{2008}]{} Qi, S., F. Y. Wang. \& Lu, T. 2008b, A\&A, 487, 853
\bibitem[\protect\citeauthoryear{}{2009}]{} Olivares, F. et al. 2009, GCN, 9215
\bibitem[\protect\citeauthoryear{}{2007}]{} Rapetti, D. et al. 2007, MNRAS, 375, 1510
\bibitem[\protect\citeauthoryear{}{1998}]{} Riess, A.G. et al. 1998, AJ, 116, 1009
\bibitem[\protect\citeauthoryear{}{2004}]{} Riess, A.G. et al. 2004, ApJ, 607, 665
\bibitem[\protect\citeauthoryear{}{2007}]{} Riess, A. G. et al. 2007, ApJ, 659, 98
\bibitem[\protect\citeauthoryear{}{2009}]{} Riess, A. G. et al. 2009, ApJ, 699, 539
\bibitem[\protect\citeauthoryear{}{2003}]{} Sahni, V. et al. 2003, JETP Lett. 77, 201
\bibitem[\protect\citeauthoryear{}{2007}]{} Schaefer, B. E. 2007, ApJ, 660, 16
\bibitem[\protect\citeauthoryear{}{2003}]{} Silva, P.T. \& Bertolami O. 2003, ApJ. 599, 829.
\bibitem[\protect\citeauthoryear{}{2003}]{} Spergel. D. N. et al. 2003, ApJS, 148, 175
\bibitem[\protect\citeauthoryear{}{2007}]{} Spergel. D. N. et al. 2007, ApJS, 170, 377
\bibitem[\protect\citeauthoryear{}{2006}]{} Tegmark, M. et al. 2006, Phys. Rev. D., 74, 123507
\bibitem[\protect\citeauthoryear{}{2004}]{} Visser, M. 2004, Class. Quant. Grav., 21, 2603
\bibitem[\protect\citeauthoryear{}{2003}]{} Vollick, D. N., 2003, Phys. Rev. D, 68, 063510
\bibitem[\protect\citeauthoryear{}{2006}]{} Wang, F. Y. \& Dai, Z. G. 2006, MNRAS, 368, 371
\bibitem[\protect\citeauthoryear{}{2009}]{} Wang, F. Y., Dai, Z. G. \& Qi, S. 2009, RAA, 9, 547
\bibitem[\protect\citeauthoryear{}{2007}]{} Wang, F. Y., Dai, Z. G. \& Zhu, Z. H. 2007, ApJ, 667, 1
\bibitem[\protect\citeauthoryear{}{2003}]{} Wang, Y., Freese, K., Gondolo, P. \& Lewis, M., 2003, ApJ, 594, 25
\bibitem[\protect\citeauthoryear{}{1989}]{} Weinberg, S. 1989, Rev. Mod. Phys, 61, 1
\bibitem[\protect\citeauthoryear{}{2001}]{} Weller, J. \& Albrecht, A., 2001, Phys. Rev. Lett., 86, 1939
\bibitem[\protect\citeauthoryear{}{2007}]{} Wood-Vasey, W. M. et al. 2007, ApJ, 666, 694
\bibitem[\protect\citeauthoryear{}{2007}]{} Wright, E. L. 2007, ApJ, 664, 633
\bibitem[\protect\citeauthoryear{}{2005}]{} Xu, D., Dai, Z. G. \& Liang. E. W. 2005, ApJ, 633, 603
\end{thebibliography}
\end{document}